\def\simgt{\lower.5ex\hbox{$\; \buildrel > \over \sim \;$}}
\def\simlt{\lower.5ex\hbox{$\; \buildrel < \over \sim \;$}}

\documentclass[useAMS,usenatbib]{mn2e}

\usepackage{graphicx}

\title[The Stellar Kinematics of Edge-on Galaxy Discs]
{Structure and Kinematics of Edge-on Galaxy Discs:\\
I. Observations of the Stellar Kinematics}
\author[M. Kregel, P.~C. van der Kruit \& K.~C. Freeman]
  {M. Kregel$^1$, 
  P.~C. van~der~Kruit$^1$\thanks{E-mail: vdkruit@astro.rug.nl} 
  and K.~C. Freeman$^2$ \\
  $^1$Kapteyn Astronomical Institute, University of Groningen,
  P.O.Box 800, 9700AV Groningen, the Netherlands\\
  $^2$Research School for Astronomy \&\ Astrophysics, Mount Stromlo
  Observatory, The Australian   National University, \\
  Private Bag, Weston Creek, 2611 Canberra, Australia}
\begin{document}

\date{Accepted. Received.}

\pagerange{\pageref{firstpage}--\pageref{lastpage}} \pubyear{2004}

\label{firstpage}

\maketitle

\begin{abstract}
We present deep optical long-slit
  spectra of 17 edge-on spiral galaxies of intermediate to late
  morphological type, mostly parallel to their major axes and in a
few cases parallel to the minor axes. 
The line-of-sight stellar kinematics are
  obtained from the stellar absorption lines using the improved
  cross-correlation technique. In general, the stellar kinematics are
  regular and can be traced well into the disc-dominated region. The
  mean stellar velocity curves are far from solid-body, indicating
  that the effect of dust extinction is not large. The line-of-sight
  stellar disc velocity dispersion correlates with the galaxy maximum
  rotational velocity, but detailed modeling is necessary to
  establish whether this represents a physical relation. In four
  spirals with a boxy- or peanut-shaped bulge we are able to detect
  asymmetric velocity distributions, having a common signature with
  projected radius in the mean line-of-sight velocity and the $h_{3}$
  and $h_{4}$ curves. In two cases this kinematic asymmetry probably
  represents the `figure-of-eight' pattern synonymous of a barred
  potential. We emphasize, however, that the signatures seen in the
  $h_{3}$ and $h_{4}$ curves may also be due to the disc seen in
  projection.

This paper has been accepted by MNRAS and is available in pdf-format
at the following URL:\\

http://www.astro.rug.nl/$\sim $vdkruit/jea3/homepage/paperI.pdf

\end{abstract}


\end{document}